\documentclass{article}%
\usepackage{amsfonts}
\usepackage{amsmath}
\usepackage{amssymb}
\usepackage{graphicx}%
\providecommand{\U}[1]{\protect\rule{.1in}{.1in}}
\begin{document}

\title{The Information Geometry of Space-time\thanks{Presented at MaxEnt 2019, the
39th International Workshop on Bayesian Inference and Maximum Entropy Methods
in Science and Engineering (June 30-- July 5, 2019, Garching bei M\"{u}nchen,
Germany). }}
\author{Ariel Caticha\\{\small Department of Physics, University at Albany--SUNY, Albany, NY 12222,
USA}}
\date{}
\maketitle

\begin{abstract}
The method of maximum entropy is used to model curved physical space in terms
of points defined with a finite resolution. Such a blurred space is
automatically endowed with a metric given by information geometry. The
corresponding space-time is such that the geometry of any embedded spacelike
surface is given by its information geometry.

The dynamics of blurred space, its \emph{geometrodynamics}, is constructed by
requiring that as space undergoes the deformations associated with evolution
in local time, it sweeps a four-dimensional space-time. This reproduces
Einstein's equations for vacuum gravity. We conclude with brief comments on
some of the peculiar properties of blurred space: There is a minimum length
and blurred points have a finite volume. There is a relativistic
\textquotedblleft blur dilation\textquotedblright. The volume of space is a
measure of its entropy.

\end{abstract}

\section{Introduction}

The problem of reconciling quantum theory (QT) and general relativity (GR) has
most commonly been addressed by preserving the framework of QT essentially
unchanged while modifying the structure and dynamics of space-time. This is
not unreasonable. Einstein's equation, $G_{\mu\nu}=8\pi G\,T_{\mu\nu}$,
relates geometry on the left to matter on the right. Since our best theories
for the matter right hand side are QTs it is natural to try to construct a
theory in which the geometrical left hand side is also of quantum mechanical
origin.\footnote{For an introduction to the extensive literature on canonical
quantization of gravity, loop quantum gravity, string theory, and causal sets
see \emph{e.g.}, \cite{Kiefer 2007}\cite{Ashtekar et al 2015}.}

Further thought however shows that this move carries a considerable risk,
particularly because the old process of quantization involves ad hoc rules
which, however successful in the past, have led to conceptual difficulties
that would immediately spread and also infect the gravitational field. One
example is the old quantum measurement problem and its closely related cousin
the problem of macroscopic superpositions. Do quantum superpositions of
space-times even make sense? In what direction would the future be? Another
example is the cosmological constant problem. Does the zero point energy of
quantum fields gravitate? Why does it not give rise to unacceptably large
space-time curvatures? Considerations such as these suggest that the issue of
whether and how to quantize gravity hinges on a deeper understanding of the
foundations of QT and also on a deeper understanding of GR and of geometry
itself --- what, after all, is \emph{distance}? Why are QT and GR framed in
such different languages? Recent developments indicate that they might be
closer than previously thought --- the link is entropy. Indeed, in the
entropic dynamics approach \cite{Caticha 2017}-\cite{Ipek Caticha 2019} QT is
derived as an application of entropic methods of inference \cite{Caticha 2012}
with a central role assigned to concepts of information geometry.\footnote{The
subject of information geometry was introduced in statistics by Fisher
\cite{Fisher 1925} and Rao \cite{Rao 1945} with important later contributions
by other authors \cite{Amari 1985}-\cite{Ay et al 2017}. Important aspects
were also independently discovered in thermodynamics \cite{Weinhold
1975}\cite{Ruppeiner 1979}.} And, on the GR side, the link between gravity and
entropy has been recognized from the early work of Bekenstein and Hawking and
further reaffirmed in more recent thermodynamic approaches to GR
\cite{Bekenstein 1973}-\cite{Jacobson 2016}.

In a previous paper \cite{Caticha 2015} we used the method of maximum entropy
to construct a model of physical space in which points are blurred; they are
defined with a finite resolution. Such a blurred space is a statistical
manifold and therefore it is automatically endowed with a Riemannian metric
given by information geometry. Our goal here is to further close the gap
between QT and GR by formulating the corresponding Lorentzian geometry of space-time.

The extension from space to space-time is not just a simple matter of applying
information geometry to four dimensions rather than three. The problem is that
information geometry leads to metrics that are positive --- statistical
manifolds are inevitably Riemannian --- which cannot reproduce the light-cone
structure of space-time. Some additional ingredient is needed. We do not model
space-time as a statistical manifold. Instead, space-time is modelled as a
four-dimensional manifold such that the geometry of all space-like embedded
surfaces is given by information geometry. We find that in the limit of a flat
space-time our model coincides with a stochastic model of space-time proposed
long ago by Ingraham by following a very different line of argument
\cite{Ingraham 1964}.

Blurred space is a curious hybrid: some features are typical of discrete
spaces while other features are typical of continuous manifolds.\footnote{It
is possible that there is some connection with the ideas formulated in the
language of spectral geometry proposed by Kempf \cite{Kempf 2009}. This is a
topic for future research.} For example, there is a minimum length and blurred
points have a finite volume. The volume of a region of space is a measure of
the number points within it, and it is also a measure of its bulk entropy.
Under Lorentz transformations the minimum length suffers a dilation which is
more analogous to the relativistic time dilation than to the familiar length contraction.

The dynamics of blurred space, its \emph{geometrodynamics}, is constructed by
requiring that as three-dimensional space undergoes the deformations
associated with time evolution it sweeps a four-dimensional space-time. As
shown in a remarkable paper by Hojman, Kucha\u{r}, and Teitelboim \cite{Hojman
et al 1976} in the context of the familiar sharp space-time this requirement
is sufficient to determine the dynamics. Exactly the same argument can be
deployed here. The result is that in the absence of matter the
geometrodynamics of four-dimensional blurred space-time is given by Einstein's
equations. The coupling of gravity to matter will not be addressed in this work.

\section{The information geometry of blurred space}

To set the stage we recall the model of blurred space as a smooth
three-dimensional manifold $\mathbf{X}$ the points of which are defined with a
finite resolution \cite{Caticha 2015}. It is noteworthy that, unlike the very
rough space-time foams expected in some models of quantum gravity, one expects
blurred space to be very smooth because irregularities at scales smaller than
the local uncertainty are suppressed. Blurriness is implemented as follows:
when we say that a test particle is located at $x\in\mathbf{X}$ (with
coordinates $x^{a}$, $a=1,2,3$) it turns out that it is actually located at
some unknown neighboring $x^{\prime}$. The probability that $x^{\prime}$ lies
within $d^{3}x^{\prime}$ is $p(x^{\prime}|x)d^{3}x^{\prime}$. Since to each
point $x\in\mathbf{X}$ one associates a distribution $p(x^{\prime}|x)$ the
space $\mathbf{X}$ is a statistical manifold automatically endowed with a
metric. Indeed, when points are blurred one cannot fully distinguish the point
at $x$ described by the distribution $p(x^{\prime}|x)$ from another point at
$x+dx$ described by $p(x^{\prime}|x+dx)$. The quantitative measure of
distinguishability \cite{Caticha 2012}\cite{Amari 1985} is the
\emph{information distance},
\begin{equation}
d\ell^{2}=g_{ab}\,(x)dx^{{}a}dx^{{}b}\,\,, \label{info metric a}%
\end{equation}
where the metric tensor $g_{ab}$ --- \emph{the information metric} --- is
given by,
\begin{equation}
g_{ab}\,(x)=\int dx^{\prime}\,p(x^{\prime}|x)\,\partial_{a}\log p(x^{\prime
}|x)\,\partial_{b}\log p(x^{\prime}|x)~.\,\, \label{info metric b}%
\end{equation}
(We adopt the standard notation $\partial_{a}=\partial/\partial x^{a}$ and
$dx^{\prime}=d^{3}x^{\prime}$.) Thus, in a blurred space \emph{distance is
distinguishability}.

In Section \ref{Discussion} we will briefly address the physical/geometrical
interpretation of $d\ell$. For now we merely state \cite{Caticha 2015} that
$d\ell$ measures the distance between two neighboring points in units of the
local uncertainty defined by the distribution $p(x^{\prime}|x)$, that is,
information length is measured in units of the local blur.

In order to completely define the information geometry of $\mathbf{X}$ which
will allow us to introduce notions of parallel transport, curvature, and so
on, one must specify a connection or covariant derivative $\nabla$. The
natural choice is the Levi-Civita connection, defined so that $\nabla
_{a}g_{bc}=0$. Indeed, as argued in \cite{Brody Hughston 1997}, the
Levi-Civita connection is to be preferred because, unlike the other $\alpha
$-connections \cite{Amari 1985}, it does not require imposing any additional
structure on the Hilbert space of functions $(p)^{1/2}$.

The next step is to use the method of maximum entropy to assign the blur
distribution $p(x^{\prime}|x)$. The challenge is to identify the constraints
that capture the physically relevant information. One might be tempted to
consider imposing constraints on the expected values of $\langle x^{\prime}%
{}^{a}-x^{a}\rangle$ and $\left\langle (x^{\prime a}-x^{a})(x^{\prime}{}%
^{b}-x^{b})\right\rangle $ but this does not work because in a curved space
neither of these constraints is covariant. This technical difficulty is evaded
by maximizing entropy on the flat space $\mathbf{T}_{P}$ that is tangent to
$\mathbf{X}$ at $P$ and then using the \emph{exponential map} (see
\cite{Caticha 2015}) to \textquotedblleft project\textquotedblright\ the
distribution from the flat $\mathbf{T}_{P}$ to the curved space $\mathbf{X}$.
It is important to emphasize that the validity of this construction rests on
the assumption that the normal neighborhood of every point $x$ --- the region
about $x$ where the exponential map is 1-1 --- is sufficiently large. The
assumption is justified provided the scale of the blur is much smaller than
the scale over which curvature effects are appreciable.

Consider a point $P\in\mathbf{X}$ with generic coordinates $x^{a}$ and a
positive definite tensor field $\gamma^{ab}(x)$. The components of
$y\in\mathbf{T}_{P}$ are $y^{a}$. The distribution $\hat{p}(y|P)$ on
$\mathbf{T}_{P}$ is assigned on the basis of information about the expectation
$\langle y^{a}\rangle_{P}$ and the variance-covariance matrix $\left\langle
y^{a}y^{b}\right\rangle _{P}$,
\begin{equation}
\langle y^{a}\rangle_{P}=0\quad\text{and}\quad\left\langle y^{a}%
y^{b}\right\rangle _{P}=\gamma^{ab}(P)~. \label{constr a}%
\end{equation}
On $\mathbf{X}$ it is always possible to transform to new coordinates%
\begin{equation}
x^{i}=X^{i}(x^{a})~, \label{NC}%
\end{equation}
such that%
\begin{equation}
\quad\gamma^{ij}(P)=\delta^{ij}\quad\text{and}\quad\partial_{k}\gamma
^{ij}(P)=0~, \label{NC a}%
\end{equation}
where $i,j,\ldots=1,2,3$. If $\gamma^{ab}$ were a metric tensor the new
coordinates would be called Riemann Normal Coordinates at $P$ (RNC$_{P}$). The
new components of $y$ are
\begin{equation}
y^{i}=X_{a}^{i}y^{a}\quad\text{where}\quad X_{a}^{i}=\frac{\partial x^{i}%
}{\partial x^{a}}~, \label{NC b}%
\end{equation}
and the constraints (\ref{constr a}) take the simpler form,
\begin{equation}
\langle y^{i}\rangle_{P}=0\quad\text{and}\quad\left\langle y^{i}%
y^{j}\right\rangle _{P}=\delta^{ij}~. \label{constr b}%
\end{equation}
We can now maximize the entropy
\begin{equation}
S[\hat{p},q]=-\int d^{3}y\,\hat{p}(y|P)\log\frac{\hat{p}(y|P)}{\hat{q}(y)}\,
\end{equation}
relative to the measure $\hat{q}(y)$ subject to (\ref{constr b}) and
normalization. Since $\mathbf{T}_{P}$ is flat we can take $\hat{q}(y)$ to be
constant and we may ignore it. The result in RNC$_{P}$ is
\begin{equation}
\hat{p}(y^{i}|P)=\frac{1}{(2\pi)^{3/2}}\,\exp\left[  -\frac{1}{2}\delta
_{ij}y^{i}y^{j}\right]  ~. \label{NC Gaussian a}%
\end{equation}
Using the inverse of eq.(\ref{NC b}) we can transform back to the original
coordinates $y^{a}$,
\begin{equation}
y^{a}=X_{i}^{a}y^{i}\quad\text{and}\quad\gamma_{ab}=X_{a}^{i}X_{b}^{j}%
\delta_{ij}~. \label{gamma ab}%
\end{equation}
The resulting distribution is also Gaussian,
\begin{equation}
\hat{p}(y^{a}|P)=\frac{(\det\gamma_{ab})^{1/2}}{(2\pi)^{3/2}}\,\exp\left[
-\frac{1}{2}\gamma_{ab}y^{a}y^{b}\right]  ~, \label{GC Gaussian a}%
\end{equation}
and the matrix $\gamma_{ab}$ of Lagrange multipliers turns out to be the
inverse of the correlation matrix $\gamma^{ab}$, $\gamma_{ab}\gamma
^{bc}=\delta_{a}^{c}$.

Next we use the exponential map to project $y^{i}$ coordinates on the flat
$\mathbf{T}_{P}$ to the RNC$_{P}$ coordinates on the curved $\mathbf{X}$,
\begin{equation}
x^{\prime i}=x^{i}(P)+y^{i}~. \label{NC x'}%
\end{equation}
The corresponding distribution $p(x^{\prime i}|P)$ induced on $\mathbf{X}$ by
$\hat{p}(y^{i}|P)$ on $\mathbf{T}_{P}$ is
\begin{equation}
p(x^{\prime i}|P)d^{3}x^{\prime}=\hat{p}(y^{i}|P)d^{3}y~,
\end{equation}
or
\begin{equation}
p(x^{\prime i}|x^{i})=\frac{1}{(2\pi)^{3/2}}\,\exp\left[  -\frac{1}{2}%
\delta_{ij}(x^{\prime i}-x^{i})(x^{\prime j}-x^{j})\right]  ~.
\label{NC Gaussian b}%
\end{equation}
Thus, in RNC$_{P}$ the distribution $p(x^{\prime i}|x^{i})$ retains the
Gaussian form. We can now invert (\ref{NC}) and transform back to the original
generic frame of coordinates $x^{a}$ and define $p(x^{\prime a}|x^{a})$ by%
\begin{equation}
p(x^{\prime a}|x^{a})d^{3}x^{\prime a}=p(x^{\prime i}|x^{i})d^{3}x^{\prime
i}~, \label{px'x}%
\end{equation}
which is an identity between scalars and holds in all coordinate systems. In
the $x^{a}$ coordinates the distribution $p(x^{\prime a}|x^{a})$ will not, in
general, be Gaussian,
\begin{equation}
p(x^{\prime a}|x^{a})=\frac{(\det\gamma_{ab})^{1/2}}{(2\pi)^{3/2}}%
\,\exp\left[  -\frac{1}{2}\delta_{ij}\left(  X^{i}(x^{\prime a})-X^{i}%
(x^{a})\right)  \left(  X^{j}(x^{\prime a})-X^{j}(x^{a})\right)  \right]  ~.
\label{GC Gaussian b}%
\end{equation}

Finally we substitute (\ref{GC Gaussian b}) into (\ref{info metric b}) to
calculate the information metric $g_{ab}$. (The integral is easily handled in
RNC$_{P}$.) The result is deceptively simple,
\begin{equation}
g_{ab}=X_{a}^{i}X_{b}^{j}\delta_{ij}=\gamma_{ab}~. \label{GC metric}%
\end{equation}
The main result of \cite{Caticha 2015} was to show that the metric $g_{ab}$ of
a blurred space is a statistical concept that measures the \textquotedblleft
degree of distinguishability\textquotedblright\ between neighboring points.
The metric is given by the Lagrange multipliers $\gamma_{ab}$ associated to
the covariance tensor $\gamma^{ab}$ that describes the blurriness of space.

\section{Space-time and the geometrodynamics of pure gravity}

The constraint that determines the dynamics is the requirement that blurred
space be a three-dimensional spacelike \textquotedblleft
surface\textquotedblright\ embedded in four-dimensional space-time. As shown
in \cite{Hojman et al 1976} the reason this condition is so constraining is
that when evolving from an initial to a final surface every intermediate
surface must also be embeddable in the same space-time and, furthermore, the
sequence of intermediate surfaces --- the path or foliation --- is not unique.
Such a \textquotedblleft foliation invariance\textquotedblright, which amounts
to the local relativity of simultaneity, is a requirement of consistency: if
there are two alternative paths to evolve from an initial to a final state,
then the two paths must lead to the same result.

Space-time is foliated by a sequence of space-like surfaces $\left\{
\Sigma\right\}  $. Points on the surface $\Sigma$ are labeled by coordinates
$x^{a}$ ($a=1,2,3$) and space-time events are labeled by space-time
coordinates $X^{\mu}$ ($\mu=0,1,2,3$). The embedding of $\Sigma$ within
space-time is defined by four functions $X^{\mu}=X^{\mu}\left(  x\right)  $.
An infinitesimal deformation of $\Sigma$ to a neighboring $\Sigma^{\prime}$ is
specified by $X^{\mu}\left(  x\right)  \rightarrow X^{\mu}\left(  x\right)
+\delta X^{\mu}\left(  x\right)  $. The deformation vector $\delta X^{\mu}(x)$
is decomposed into normal and tangential components,
\begin{equation}
\delta X^{\mu}=\delta X^{\bot}n^{\mu}+\delta X^{a}X_{a}^{\mu}~,
\label{deformation vector}%
\end{equation}
where $n^{\mu}$ is the unit normal to the surface and the three vectors
$X_{a}^{\mu}=\partial X^{\mu}/\partial x^{a}$ are tangent to the coordinate
lines $x^{a}$ ($n_{\mu}n^{\mu}=-1$, $n_{\mu}X_{a}^{\mu}=0$).

We assume a phase space endowed with a symplectic structure: the basic
dynamical variables are the surface metric $g_{ab}(x)$ and its canonically
conjugate momentum $\pi^{ab}(x)$. This leads to a Hamiltonian dynamics where
the super-Hamiltonian $H_{\perp}(x)[g,\pi]$ and the super-momentum
$H_{a}(x)[g,\pi]$ generate normal and tangential deformations respectively. In
order for the dynamics to be consistent with the kinematics of deformations
the Poisson brackets of $H_{\bot}$ and $H_{a}$ must obey two sets of
conditions \cite{Teitelboim 1973}\cite{Kuchar 1973}. First, they must close in
the same way as the \textquotedblleft group\textquotedblright\ of
deformations, that is, they must provide a representation of the
\textquotedblleft algebra\textquotedblright\ of deformations\footnote{The
quotes in \textquotedblleft group\textquotedblright\ and \textquotedblleft
algebra\textquotedblright\ are a reminder that the set of deformations do not
form a group. The composition of two successive deformations is itself a
deformation but it depends on the surface to which the first deformation is
applied.},
\begin{align}
\lbrack H_{\bot}(x),H_{\bot}(x^{\prime})]  &  =\left(  g^{ab}(x)H_{b}%
(x)+g^{ab}(x^{\prime})H_{b}(x^{\prime})\right)  \partial_{ax}\delta
(x,x^{\prime})~,\label{PB 1}\\
\lbrack H_{a}(x),H_{\bot}(x^{\prime})]  &  =H_{\bot}(x)\partial_{ax}%
\delta(x,x^{\prime})~,\label{PB 2}\\
\lbrack H_{a}(x),H_{b}(x^{\prime})]  &  =H_{a}\,(x^{\prime})\partial_{b}%
\delta(x,x^{\prime})+H_{b}(x)\partial_{a}\delta(x,x^{\prime})~. \label{PB 3}%
\end{align}
And second, the initial values of the variables $g_{ab}$ and $\pi^{ab}$ must
be restricted to obey the weak constraints
\begin{equation}
H_{\bot}(x)\approx0\quad\text{and}\quad H_{a}(x)\approx0~.
\label{Hp Ha constr}%
\end{equation}
A remarkable feature of the resulting dynamics is that once the constraints
(\ref{Hp Ha constr}) are imposed on one initial surface $\Sigma$ they will be
satisfied automatically on all subsequent surfaces. As shown in \cite{Hojman
et al 1976} the generators that satisfy (\ref{PB 1}-\ref{PB 3}) are
\begin{align}
H_{a}  &  =-2\nabla_{b}\pi_{a}^{b}~,\label{H 1}\\
H_{\perp}  &  =2\kappa G_{abcd}\pi^{ab}\pi^{cd}-\frac{1}{2\kappa}%
g^{1/2}(R-2\Lambda)~,\\
G_{abcd}  &  =\frac{1}{2}g^{-1/2}\left(  g_{ac}g_{bd}+g_{ad}g_{bc}%
-g_{ab}g_{cd}\right)  ~, \label{H 3}%
\end{align}
where $\kappa$ and $\Lambda$ are constants which, once the coupling to matter
is introduced, can be related to Newton's constant $G=c^{4}\kappa/8\pi$ and to
the cosmological constant $\Lambda$. Equations (\ref{Hp Ha constr}-\ref{H 3})
are known to be equivalent to Einstein's equations in vacuum.

To summarize: (a) Space-time is constructed so that the geometry of any
embedded spacelike surface is given by information geometry. (b) The
geometrodynamics of blurred space is given by Einstein's equations. These are
the main conclusions of this paper.

\section{Discussion}

\label{Discussion}

\paragraph*{Dimensionless distance? ---}

As with any information geometry the distance $d\ell$ given in
eqs.(\ref{info metric a}-\ref{info metric b}) turns out to be dimensionless.
The interpretation \cite{Caticha 2015} is that an information distance is
measured distances in units of the local uncertainty --- the blur. To make
this explicit we write the distribution (\ref{NC Gaussian b}) that describes a
blurred point in RNC$_{P}$ in the form
\begin{equation}
p(x^{\prime i}|x^{i})=\frac{1}{(2\pi\ell_{0}^{2})^{3/2}}\,\exp\left[
-\frac{1}{2\ell_{0}^{2}}\delta_{ij}(x^{\prime i}-x^{i})(x^{\prime j}%
-x^{j})\right]  ~, \label{NC Gaussian c}%
\end{equation}
so that the information distance between two neighboring points is%

\begin{equation}
d\ell^{2}=\frac{1}{\ell_{0}^{2}}\delta_{ij}dx^{i}dx^{j}~. \label{dist c}%
\end{equation}
Since the blur $\ell_{0}$ is the only unit of length available to us (there
are no external rulers) it follows that $\ell_{0}=1$ but it is nevertheless
useful to write our equations showing $\ell_{0}$ explicitly. In
(\ref{NC Gaussian c}) the two points $x$ and $x^{\prime}$ are meant to be simultaneous.

\paragraph*{Minimum length ---}

To explore the geometry of blurred space it helps to distinguish the abstract
\textquotedblleft mathematical\textquotedblright\ points that are sharply
defined by the coordinates $x$ from the more \textquotedblleft
physical\textquotedblright\ blurred points. We shall call them c-points and
b-points respectively. In RNC$_{P}$ the distance between two c-points located
at $x$ and at $x+\Delta x$ is given by (\ref{dist c}). To find the
corresponding distance $\Delta\lambda$ between two b-points located at $x$ and
at $x+\Delta x$ we recall that when we say a test particle is at $x$ it is
actually located at $x^{\prime}=x+y$ so that%
\begin{equation}
\Delta\lambda^{2}=\frac{1}{\ell_{0}^{2}}\delta_{ij}(\Delta x^{i}+\Delta
y^{i})(\Delta x^{j}+\Delta y^{j})~.
\end{equation}
Taking the expectation over $y$ with the probability (\ref{NC Gaussian c}) ---
use $\left\langle y^{i}\right\rangle =0$ and $\left\langle y^{i}%
y^{j}\right\rangle =\ell_{0}^{2}\,\delta^{ij}$ --- we find
\begin{equation}
\langle\Delta\lambda^{2}\rangle=\frac{1}{\ell_{0}^{2}}\delta_{ij}%
\langle(\Delta x^{i}+\Delta y^{i})(\Delta x^{j}+\Delta y^{j})\rangle
=\Delta\ell^{2}+6~. \label{dist b}%
\end{equation}
We see that even as $\Delta x\rightarrow0$ and the two b-points coincide we
still expect a minimum \emph{rms} distance of $\sqrt{6}\ell_{0}$.

\paragraph*{Blur dilation ---}

The size of the blur of space is a length but it does not behave as the length
of a rod. When referred to a moving frame it does not undergo a Lorentz
contraction. It is more analogous to time dilation. Just as a clock marks time
by ticking along the time axis; so are lengths measured by ticking $\ell_{0}$s
along them. By the principle of relativity all inertial observers measure the
same blur in their own rest frames --- the proper blur $\ell_{0}$. Relative to
another inertial frame the blur is dilated to $\gamma\ell_{0}$ where $\gamma$
is the usual relativistic factor. This implies the proper blur $\ell_{0}$ is
indeed the minimum attainable.

\paragraph*{The volume of a blurred point: is space continuous or discrete?
---}

A b-point is smeared over the whole of space but we can still define a useful
measure of its volume by adding all volume elements $g^{1/2}(x^{\prime}%
)d^{3}x^{\prime}$ weighed by the scalar density $p(x^{\prime}|x)/g^{1/2}%
(x^{\prime})$. Therefore in $\ell_{0}$ units a blurred point has unit volume.
This means that we can measure the volume of a finite region of space by
\emph{counting} the number of b-points it contains. It also means that the
number of distinguishable b-points within a region of finite volume is finite
which is a property one would normally associate to discrete spaces. In this
sense blurred space is both continuous and discrete. (See also \cite{Kempf
2009}.)

\paragraph*{The entropy of space ---}

The statistical state of blurred space is the joint distribution of all the
$y_{x}$ variables associated to every b-point $x$. We assume that the $y_{x}$
variables at different $x$s are independent, and therefore their joint
distribution is a product,
\begin{equation}
\hat{P}[y]=%
%TCIMACRO{\tprod \limits_{x}}%
%BeginExpansion
{\textstyle\prod\limits_{x}}
%EndExpansion
\hat{p}\left(  y_{x}|x\right)  ~. \label{state space a}%
\end{equation}
From (\ref{GC Gaussian a}) and (\ref{GC metric}) the distribution $\hat
{p}\left(  y_{x}^{a}|x\right)  $ in the tangent space $\mathbf{T}_{x}$ is
Gaussian,
\begin{equation}
\hat{p}(y_{x}|x)=\frac{(\det g_{x})^{1/2}}{(2\pi)^{3/2}}\,\exp\left[
-\frac{1}{2}g_{ab}(x)y_{x}^{a}y_{x}^{b}\right]  ~, \label{state space b}%
\end{equation}
which shows explicitly how the information metric $g_{ab}$ determines the
statistical state of space.

Next we calculate the total entropy of space,
\begin{equation}
S[\hat{P},\hat{Q}]=-\int Dy\,\hat{P}[y]\log\frac{\hat{P}[y]}{\hat{Q}%
[y]}\overset{\text{def}}{=}S[g] \label{Sg a}%
\end{equation}
relative to the uniform distribution%
\begin{equation}
\hat{Q}[y|g]=%
%TCIMACRO{\tprod \nolimits_{x}}%
%BeginExpansion
{\textstyle\prod\nolimits_{x}}
%EndExpansion
g^{1/2}(x)~, \label{prior y}%
\end{equation}
which is independent of $y$ --- a constant. Since the $y$'s in
eq.(\ref{state space a}) are independent variables the entropy is additive,
$S[g]=%
%TCIMACRO{\tsum \nolimits_{x}}%
%BeginExpansion
{\textstyle\sum\nolimits_{x}}
%EndExpansion
S(x)$, and we only need to calculate the entropy $S(x)$ associated to a
b-point at a generic location $x$,
\begin{equation}
S(x)=-\int d^{3}y\,\hat{p}(y|x)\log\frac{\hat{p}(y|x)}{g^{1/2}(x)}=\frac{3}%
{2}\log2\pi e=s_{0}~.
\end{equation}
Thus, the entropy per b-point is a numerical constant $s_{0}$ and the entropy
of any region $R$ of space, $S_{R}[g]$, is just its volume,%

\begin{equation}
S_{R}[g]=%
%TCIMACRO{\dsum \nolimits_{x\in R}}%
%BeginExpansion
{\displaystyle\sum\nolimits_{x\in R}}
%EndExpansion
S(x)=s_{0}\int_{R}d^{3}x\,g^{1/2}(x)~. \label{Sg b}%
\end{equation}
Thus, the entropy of a region of space is proportional to the number of
b-points within it and is proportional to its volume.

\paragraph*{Canonical quantization of gravity? ---}

The picture of space as a smooth blurred statistical manifold stands in sharp
contrast to ideas inspired from various models of quantized gravity in which
the short distance structure of space is dominated by extreme fluctuations.
From our perspective it is not surprising that attempts to quantize gravity by
imposing commutation relations on the metric tensor $g_{ab}$ have not been
successful. The information geometry approach suggests a reason why:
quantizing the Lagrange multipliers $g_{ab}=\gamma_{ab}$ would be just as
misguided as formulating a quantum theory of fluids by imposing commutation
relations on those Lagrange multipliers like temperature, pressure, or
chemical potential, that define the thermodynamic macrostate.

\paragraph*{Physical consequences of a minimum length? ---}

A minimum length will eliminate the short wavelength divergences in QFT. This
in turn will most likely illuminate our understanding of the cosmological
constant and affect the scale dependence of running coupling constants. One
also expects that QFT effects that are mediated by short wavelength
excitations should be suppressed. For example, the lifetime of the proton
ought to be longer than predicted by grand-unified theories formulated in
Minkowski space-time. The nonlocality implicit in a minimum length might lead
to possible violations of CPT symmetry with new insights into
matter-antimatter asymmetry. Of particular interest would be the early
universe cosmology where inflation might amplify minimum-length effects
possibly making them observable.

\paragraph*{Acknowledgments}

I would like to thank N. Carrara, N. Caticha, S. Ipek, and P. Pessoa, for
valuable discussions.

\end{document}